\documentclass[3p,times,procedia]{elsarticle}
\usepackage{ecrc}
\usepackage[bookmarks=false]{hyperref}
    \hypersetup{colorlinks,
      linkcolor=blue,
      citecolor=blue,
      urlcolor=blue}

\usepackage{amssymb}

\usepackage[figuresright]{rotating}

\usepackage{multirow}
\usepackage{multicol}

\begin{document}
\pagenumbering{gobble}
\begin{frontmatter}


\title{An objective evaluation of the effects of recording conditions and speaker characteristics in multi-speaker deep neural speech synthesis}


\author[a,b]{Be\'{a}ta L\H{o}rincz\corref{cor1}} 
\author[a]{Adriana Stan}
\author[a]{Mircea Giurgiu}

\address[a]{Technical University, Communications Department,  Cluj-Napoca, Romania}
\address[b]{Babe{\textcommabelow s}-Bolyai University, Faculty of Mathematics and Computer Science, Cluj-Napoca, Romania}

\begin{abstract}
Multi-speaker spoken datasets enable the creation of text-to-speech synthesis (TTS) systems which can output several voice identities. The multi-speaker (MSPK) scenario also enables the use of fewer training samples per speaker. However, in the resulting acoustic model, not all speakers exhibit the same synthetic quality, and some of the voice identities cannot be used at all.     

In this paper we evaluate the influence of the recording conditions, speaker gender, and speaker particularities over the quality of the synthesised output of a deep neural TTS architecture, namely Tacotron2. The evaluation is possible due to the use of a large Romanian parallel spoken corpus containing over 81 hours of data.
Within this setup, we also evaluate the influence of different types of text representations: orthographic, phonetic, and phonetic extended with syllable boundaries and lexical stress markings. 

We evaluate the results of the MSPK system using the objective measures of equal error rate (EER) and word error rate (WER), and also look into the distances between natural and synthesised t-SNE projections of the embeddings computed by an accurate speaker verification network. 
The results show that there is indeed a large correlation between the recording conditions and the speaker's synthetic voice quality. The speaker gender does not influence the output, and that extending the input text representation with syllable boundaries and lexical stress information does not equally enhance the generated audio across all speaker identities. The visualisation of the t-SNE projections of the natural and synthesised speaker embeddings show that the acoustic model shifts some of the speakers' neural representation, but not all of them. As a result, these speakers have lower performances of the output speech. 

\end{abstract}

\cortext[cor1]{Corresponding author}
\end{frontmatter}

\email{beata.lorincz@ubbcluj.ro}




\section{Introduction}
\label{intro}

Text-to-speech (TTS) synthesis systems have been widely studied in the research community with an increased interest since the appearance of the deep learning based solutions. The deep neural network (DNN) based speech synthesis achieved a naturalness close to that of human speech~\cite{shen2018natural}, making it suitable for commercial applications as well.
These TTS systems are now extensively used in our daily lives for virtual assistants, navigation systems or the generation of audio content for any written text.  
While most TTS systems use a single voice identity, speech corpora consisting of recordings belonging to multiple speakers allows the training of the so called \emph{multi-speaker speech synthesis systems (MSPK)}. This synthesis setup has the advantage of integrating multiple voice identities into a single acoustic model. 
Some of the first approaches to multi-speaker speech synthesis were based on Hidden Markov Models (HMM) and were built upon average voice models trained on data from multiple speakers and were then adapted to new speaker identities~\cite{yamagishi2009robust}. This methodology requires separate acoustic models, and the quality of the output speech is rather poor~\cite{watts2016hmms}, even though there are methods that aim to improve the quality degradation in HMM-based speech synthesis~\cite{biagetti2018hmm}.
The development of systems supporting multiple voice identities in deep neural synthesis was first approached by adapting the neural network architecture fully or partially to new target speakers~\cite{kons2019high}. Other studies have proposed the training of a speaker encoder network trained jointly with the TTS model~\cite{chen2018sample}. More recently, the speaker identity is most commonly represented with the help of the so-called~\emph{speaker embeddings} that are learnt during training~\cite{cooper2020zero}. These representations of the speaker characteristics learn the relevant speaker information for the task of speech synthesis~\cite{gibiansky2017deep}. Precursors for these representations were the fixed embeddings, such as i-vectors~\cite{cardinal2015speaker} and x-vectors~\cite{snyder2018x}. Speaker embeddings are advantageous as the number of parameters in the synthesis system are reduced, and therefore require less data from each speaker. Several studies analyse the utility of speaker embeddings: in~\cite{cooper2020zero} different types of speaker embeddings are explored and their feasibility for modelling speaker identity is examined. In~\cite{mitsui2020multi} the authors report that speaker representations can be learnt by latent variable models of deep Gaussian processes, producing representations that efficiently learn the speaker similarities or dissimilarities. 
Multi-speaker speech synthesis systems implementing speaker embeddings are presented in \cite{gibiansky2017deep, park2019multi, chen2020multispeech} with some of the models including very large number of speakers~\cite{ping2018deep}.

Other approaches to multi-speaker speech synthesis are meta-learning methods or transfer learning. \cite{jia2018transfer}~employs a speaker verification network to train the speaker encoder for multi-speaker speech synthesis. The authors of \cite{chen2018sample} propose a meta-learning method by training a fast-learner network that can be adapted rapidly to new speakers.

Neural speech synthesis, and specifically multi-speaker synthesis requires large amounts of speech data. Corpora that contain multiple voices are usually composed of data of different styles, various recording conditions and hence of varying quality. Evaluating and selecting subsets of the available data has shown to result in better quality speech output compared with models trained on larger sets of data \cite{gallegos2020unsupervised}. 

Starting from this overview we exploit a Romanian parallel corpus that contains speech recorded under different conditions, and evaluate if particular voice characteristics or recording conditions influence the resulting quality of the respective voice identity in a multi-speaker TTS scenario. We also analyse if different text representations fed to the neural TTS system boost or deteriorate the synthesised speech of the different speakers. 
The main contributions of our work are: 1) an objective evaluation of speaker characteristics' influence in multi-speaker neural speech synthesis; and 2) an evaluation of three different types of text representations fed as input to the neural network.  

The paper is organized as follows: Section~\ref{method} describes the neural TTS system, the speech corpus and text representations used in the evaluation. Section~\ref{results} shows and discusses the results, while conclusions are drawn in Section~\ref{conclusions}.

\section{Evaluation setup}
\label{method}


The Tacotron2 TTS system is currently one of the most natural synthesis systems, with a reported MOS score of 4.53~\cite{shen2018natural}. Its main advantages are drawn from the use of recurrent structures within the neural architecture and a multi-head attention. However, the recurrence also makes it unstable in generating long output sequences and increases the inference speed. Yet, if we want to evaluate the influence of speaker particularities over the performance of a TTS system, it is essential to use an architecture which can--hypothetically--reproduce the naturalness and speaker identity as accurate as possible. We also employ the use of a large parallel spoken corpus consisting of over 81 hours of data collected from 47 speakers. Because the text representation has also been found to affect the quality of the synthesised output~\cite{peiro2020naturalness}, we look into different linguistic information present at the input of the TTS data. All three components of our evaluation are presented into more detail in the following subsections.


The implementation of the multi-speaker TTS system relies on the architecture of Tacotron2~\cite{shen2018natural} neural TTS system. 
We extended the NVIDIA single-speaker implementation\footnote{https://github.com/NVIDIA/tacotron2} 
to support multi-speaker speech synthesis inspired by the Mellotron \cite{valle2020mellotron} implementation.\footnote{https://github.com/NVIDIA/mellotron}
In this architecture, the 
speaker embeddings are appended to the text encoder input, and the resulting hidden features are then used to condition the audio decoder, yielding the output Mel spectrogram.  
The speaker embeddings are 128-dimensional and are learnt during the training process. The multi-speaker model was trained for 250k steps with a learning rate of $10^{-3}$ on batches of size 16.

The conversion of the Mel spectrogram into the audio waveform uses the Waveglow~\cite{prenger2019waveglow} flow-based neural vocoder. The model was pre-trained on a large number of English speakers,\footnote{https://github.com/NVIDIA/waveglow} but did not include the SWARA and SWARA 2.0 speakers.

\subsection{Speech corpus}

High-quality speech corpora is essential in neural speech synthesis. In this work we start from the large parallel Romanian dataset called \textbf{SWARA}~\cite{stan2017swara}. SWARA contains 17 volunteer speakers each reading aloud between 1000 and 1500 utterances (the same across all speakers) in a controlled studio environment. The total duration of the recordings set is around 21 hours, and was previously successfully applied to train various TTS systems. 

The initial SWARA dataset was recently extended with 29 new speakers reading the same utterances, as well as an additional short story. The average number of utterances pertaining to each speaker is 1747, and the total duration of the recordings for all speakers amounts to 59 hours and 39 minutes. Due to the current global pandemic situation, the recordings were performed in the speakers' home environments using the RECOApy tool~\cite{stan2020recoapy}, and were lightly checked for errors by the authors. Some of the issues notices in the recordings refer to the reverberation and background noise presence, as well as some utterances which are chopped either in the beginning or in the end, thus yielding incorrect text-to-audio alignments. 
We refer to this new set of recordings as \textbf{SWARA 2.0}.    
All utterances in SWARA and SWARA 2.0 are sampled at 48kHz with 16bps, and were downsampled to 22kHz for the Tacotron2 training.  

For the multi-speaker neural TTS evaluation we selected a set of 500 sentences uttered by 37 of the 47 available speakers. The total duration of training data used from the selected speakers tallies up to 28 hours and 57 minutes. An additional set of 12 utterances per speaker were set aside for testing purposes. There are 22 female and 15 male voices\footnote{We note that the male/female classification was performed based on the biological gender of the speakers, and did not require a self-assessment on behalf of the speakers.} in the selected set. Out of the total, 9 females and 7 males are part of the SWARA corpus, and the rest pertain to SWARA 2.0. The total duration of the selected speech dataset is approximately 30 hours, with an average of 50 minutes of speech per speaker. Each speaker is given a unique ID which does not infer their identity.

\subsection{Text representation}

\begin{figure}[t!]
\scriptsize
\includegraphics[width=0.45\textwidth]{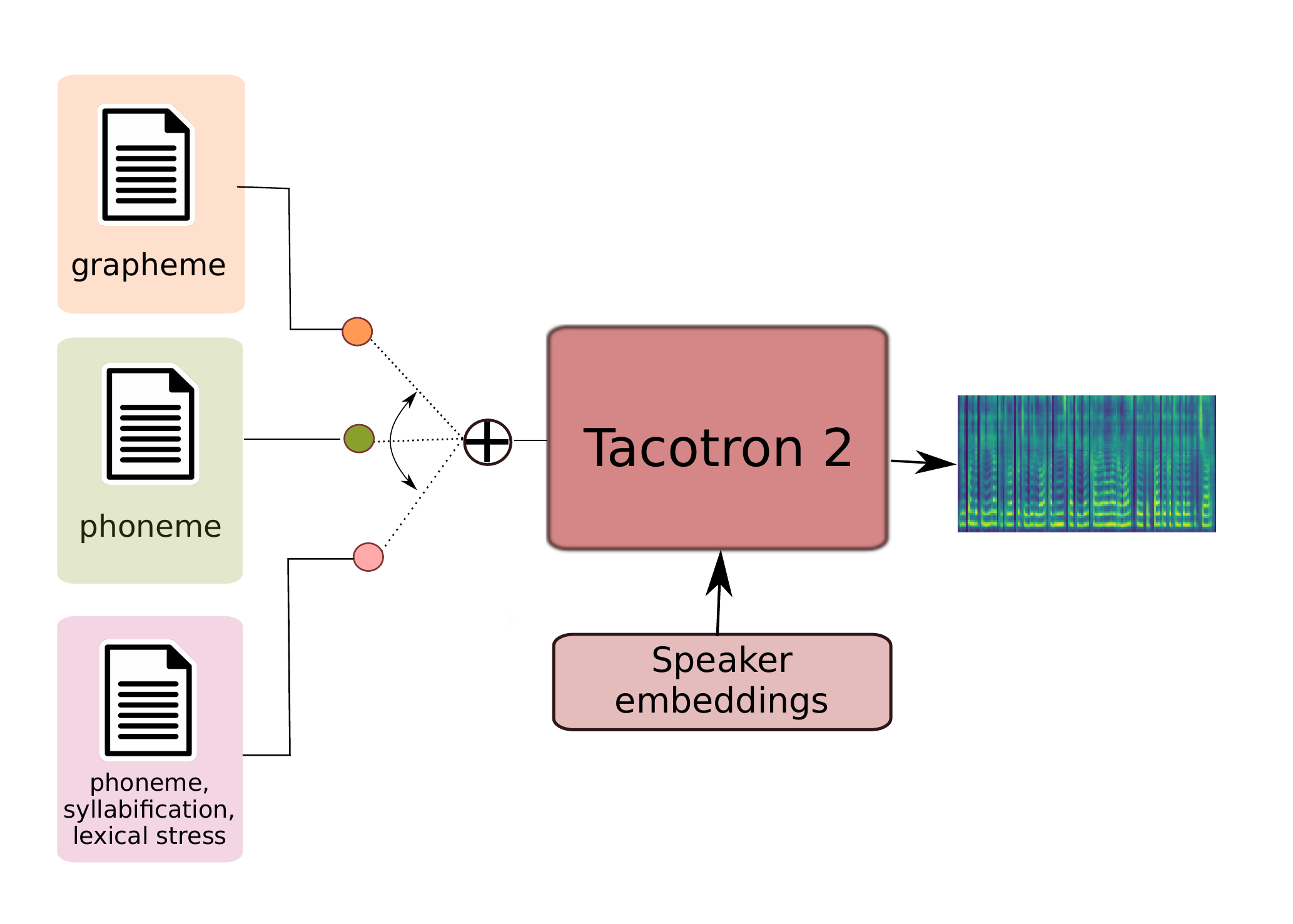}
\centering
\caption{Overview of the TTS system flow: the input text is fed in one of the three representations: graphemes, phonemes or phonemes with syllabification and lexical stress. The speaker embeddings are appended to Tacotron2's text encoder and condition the audio decoder module.}
\label{fig:tts_system_flow}
\end{figure}

The early versions of end-to-end neural network TTS systems relied mostly on grapheme representations~\cite{sotelo2017char2wav, wang2017tacotron}. However, this representation is not suitable for phonetically rich languages~\cite{vythelingum2018acoustic}, and especially for the alignment component of sequence-to-sequence architectures. Moving from graphemes to phonemic representations, and even augmenting the data with additional features, such as lexical stress markings, syllable boundaries or part-of-speech tagging~\cite{peiro2020naturalness, taylor2020enhancing} enhances the naturalness of the speech output. 
We also test this hypothesis in the context of the multi-speaker Tacotron2 system, and use three types of text representation as input for the TTS: i) orthographic/graphemic representation \textbf{GR}; ii) phonemic representation \textbf{PH}; and iii) phonemic representation extended with the syllable boundary and lexical stress markings \textbf{EXT}. 
Examples of the text representations for the three scenarios are presented in Table~\ref{tab:text_inputs}. A simple diagram of the general architecture of our TTS system is represented in~Figure~\ref{fig:tts_system_flow}.

\begin{table}[h]
\caption{Text representation examples used for the multi-speaker TTS. Syllable boundaries are marked with a hyphen, and the stress is marked with an apostrophe symbol. The phonemic representations use the SAMPA notation.}
\label{tab:text_inputs}
\centering
\begin{tabular}{p{0.2\linewidth} p{0.1\linewidth} p{0.6\linewidth}}
\toprule
\textbf{Text representation} & \textbf{ID} & \textbf{Example} \\
\colrule
graphemes & \textbf{GR} & Acesta se referă însă doar la proprietățile din capitală. \\
phonemes & \textbf{PH} & atSesta se refer@ 1ns@ do\_Xar la propriet@tsile din kapital@. \\
phonemes + syllabification + lexical stress & \textbf{EXT} & a-tSe's-ta se re-fe'-r@ 1'n-s@ do\_Xa'r la pro-pri-e-t@'-tsi-le din ka-pi-ta'-l@. \\
\botrule
\end{tabular}
\end{table}

To obtain the phonemic and extended text representations, we rely on a Transformed-based model trained on the RoLEX dataset~\cite{rolex}. RoLEX contains over 330,000 manual entries including information about lemma, part-of-speech tagging, syllabification, lexical stress and phonemic transcription. The Transformer model trained on RoLEX achieved a word error rate of 3.08\%, where a large number of errors were a result of the incorrect lexical stress assignment.  
This model is a follow-up work for the concurrent prediction task of the same lexical features using recurrent and convolutional architectures \cite{lHorincz2020concurrent}. 


\section{Evaluation results}
\label{results}



A first evaluation of the synthesised spoken output was performed using two objective measures: equal error rate (EER) pertaining to speaker similarity, and word error rate (WER) pertaining to speech intelligibility. For both measures we synthesised the 12 test utterances from each speaker within the dataset using the three text input representations for the TTS systems. Audio samples for selected voices are available on our website: \url{https://speech.utcluj.ro/multispeaker\_tts\_kes2021/}.
The EER measure was computed by employing the speaker verification (SV) network described in~\cite{chung2020in}.\footnote{We used the implementation and pre-trained models available at: https://github.com/clovaai/voxceleb\_trainer.} 
This network was trained on 5994 speakers from the Voxceleb dataset~\cite{nagrani2017voxceleb} and reports an EER of 2.21\% for the best performing model.
In the EER evaluation we paired the 12 synthesised test utterances from each speaker with natural counterparts from the same speaker and also from the other speakers. This way we obtained a set of 888 pairs of natural vs synthesised neural embedding comparisons. The resulting EER values for each speaker are summarized in Table~\ref{tab:EER_values}. The table groups the studio vs home recording conditions, and male vs female identity, and also reports the values obtained for the different text representations used as input to the Tacotron2 network. Best results in each column are highlighted in boldface. 

The WER calculations used the same set of 12 synthesised sentences per system from each speaker. The synthesised sentences were transcribed using a general purpose high quality automatic speech recognition tool for Romanian~\cite{georgescu2019kaldi}. The ASR's WER on the natural test samples is 10.45\%.
The WER values for each speaker, recording condition, speaker's gender and text input representation are shown in Table~\ref{tab:WER_values}.

\begin{table}[t!]
\caption{\textbf{EER [\%]} values for each speaker grouped by recording conditions and gender for each of the three text representations used as input for the TTS system: \textbf{GR} - graphemes; \textbf{PH} - phonemes; and \textbf{EXT} - phonemes plus syllabification and lexical stress markings. The best performing speakers in each column are highlighted with boldface.}
\label{tab:EER_values}
\centering
\scriptsize
\begin{tabular}{cccc|cccc|cccc|cccc}
\toprule
\multicolumn{8}{c}{\textbf{Studio recording}} \vline & 
\multicolumn{8}{c}{\textbf{Home recording}} \\
\hline
\multicolumn{4}{c}{\textbf{Female}} \vline & \multicolumn{4}{c}{\textbf{Male}} \vline &
\multicolumn{4}{c}{\textbf{Female}} \vline & \multicolumn{4}{c}{\textbf{Male}} \\
\colrule
\textbf{Speaker} & \textbf{GR} & \textbf{PH} & \textbf{EXT} & 
\textbf{Speaker} & \textbf{GR} & \textbf{PH} & \textbf{EXT} &
\textbf{Speaker} & \textbf{GR} & \textbf{PH} & \textbf{EXT} &
\textbf{Speaker} & \textbf{GR} & \textbf{PH} & \textbf{EXT} \\
\colrule
BAS & 16.66 & 25.00 & 16.66 & FDS & 16.66 & 16.66 & 8.33 &
BGL & 16.66 & 16.66 & 16.66 & BIM & \textbf{8.33} &\textbf{8.33} & 16.66 \\

BEA & 8.33 & 16.66 & \textbf{0.00} & PCS & \textbf{0.00} & 16.66 & \textbf{0.00}&
BMM & \textbf{0.00} & 8.33 & 16.66 & BVL & 50.00 & 41.66 & 41.66 \\

DCS & 16.66 & 25.00 & 16.66 & PSS & 8.33 & \textbf{0.00} & \textbf{0.00}  &
CCL & 8.33 & 8.33 & \textbf{0.00} & MGL & 16.66 & 16.66 & 16.66 \\

DDM & 8.33 & 8.33 & 16.66 & RMS & \textbf{0.00} & \textbf{0.00} & \textbf{0.00}&
CMM & 41.66 & 41.66 & 41.66 & NLL & 16.66 & 16.66 & \textbf{0.00} \\

EME & 8.33 & 8.33 & 8.33 & SDS & 8.33 & \textbf{0.00} & 16.66 &
DOL & 33.33 & 16.66 & 16.66 & PDL & \textbf{8.33} & 16.66 & 25.00 \\

HTM & 8.33 & 8.33 & 8.33 & SGS & 8.33 & \textbf{0.00} & 16.66  &
GAM & 50.00 & 58.33 & 58.33 & PTL & 25.00 & 25.00 & 16.66 \\

MAR & 8.33 & 16.66 & \textbf{0.00} & TSS & 16.66 & 16.66 & 8.33 &
GIM & 16.66 & 8.33 & 16.66  & SRL & 25.00 & 25.00 & 16.66 \\

PMM & 16.66 & 16.66 & 16.66 & & & & &
GNM & 16.66 & 16.66 & 16.66 & ZPL & 16.66 & \textbf{8.33} & 16.66 \\

SAM & \textbf{0.00} & \textbf{0.00} & \textbf{0.00} & & & & &
MAL & 16.66 & 25.00 & 25.00 & & & & \\

& & & & & & & &
MRL & 33.33 & 41.66 & 33.33 & & & & \\

& & & & & & & &
OGL & \textbf{0.00} & \textbf{0.00} & \textbf{0.00} & & & & \\

& & & & & & & &
PBL & 16.66 & 16.66 & 16.66 & & & & \\

& & & & & & & &
SMM & 16.66 & \textbf{0.00} & \textbf{0.00} & & & & \\ \hline


\botrule
\end{tabular}
\end{table}

\begin{table}[t!]
\caption{\textbf{WER [\%]} values for each speaker grouped by recording conditions and gender for each of the three text representations used as input for the TTS system: \textbf{GR} - graphemes; \textbf{PH} - phonemes; and \textbf{EXT} - phonemes plus syllabification and lexical stress markings. The best performing speaker in each column is highlighted with boldface.}
\label{tab:WER_values}
\centering
\scriptsize
\begin{tabular}{cccc|cccc|cccc|cccc}
\toprule
\multicolumn{8}{c}{\textbf{Studio recording}} \vline & 
\multicolumn{8}{c}{\textbf{Home recording}} \\
\hline
\multicolumn{4}{c}{\textbf{Female}} \vline & \multicolumn{4}{c}{\textbf{Male}} \vline &
\multicolumn{4}{c}{\textbf{Female}} \vline & \multicolumn{4}{c}{\textbf{Male}} \\
\colrule
\textbf{Speaker} & \textbf{GR} & \textbf{PH} & \textbf{EXT} & 
\textbf{Speaker} & \textbf{GR} & \textbf{PH} & \textbf{EXT} &
\textbf{Speaker} & \textbf{GR} & \textbf{PH} & \textbf{EXT} &
\textbf{Speaker} & \textbf{GR} & \textbf{PH} & \textbf{EXT} \\
\colrule
BAS & 19.67 & 19.76 & 31.71 & FDS & 16.98 & 16.86 & 16.26 &
BGL & 41.45 & 51.99 & 42.40 & BIM & 18.69 & \textbf{14.99} & 20.57 \\

BEA & 12.94 & 16.84 & \textbf{10.13} & PCS & 36.17 & 28.47 & 30.55  &
BMM & 20.82 & 20.68 & 25.29 & BVL & 83.96 & 89.23 & 98.80 \\

DCS & 13.75 & 21.47 & 14.76 & PSS & 14.84 & 14.64 & 20.20&
CCL & 30.32 & 26.22 & 33.86 & MGL & 16.52 & 20.73 & \textbf{16.10} \\

DDM & 15.36 & 16.42 & 15.59 & RMS & \textbf{10.61} & \textbf{12.21} & \textbf{10.13} &
CMM & 20.54 & 23.78 & 20.54 & NLL & \textbf{14.76} & 18.81 & 17.10 \\
EME & 14.32 & 13.14 & 17.22 & SDS & 22.05 & 13.83 & 21.93  &
DOL & \textbf{19.82} & \textbf{12.81} & \textbf{11.05} & PDL & 54.41 & 37.85 & 54.00 \\

HTM & 12.56 & 19.66 & 24.64 & SGS & 29.95 & 20.19 & 21.47&
GAM & 30.19 & 34.72 & 38.69 & PTL & 25.78 & 18.00 & 34.96 \\

MAR & \textbf{8.85} & 12.67 & 16.21 & TSS & 20.91 & 14.29 & 14.76 &
GIM & 34.74 & 30.31 & 22.02 & SRL & 46.64 & 26.64 & 42.40 \\

PMM & 11.40 & \textbf{10.96} & 17.91 & & & & &
GNM & 31.23 & 30.21 & 33.58 & ZPL & 26.15 & 27.30 & 26.24 \\

SAM & 9.69 & 11.05 & 17.07 & & & & &
MAL & 20.89 & 19.16 & 18.00 & & & & \\

& & & & & & & &
MRL & 37.21 & 33.55 & 14.87 & & & & \\

& & & & & & & &
OGL & 26.54 & 17.27 & 30.38 & & & & \\

& & & & & & & &
PBL & 67.93 & 71.94 & 42.84 & & & & \\

& & & & & & & &
SMM & 23.47 & 21.40 & 30.75 & & & & \\ \hline

\botrule
\end{tabular}
\end{table}

\begin{table}[t]
\caption{Average \textbf{EER[\%]} and \textbf{WER[\%]} values across the different evaluation criteria.}
\label{tab:EER_WER_avg}
\centering
\scriptsize
\begin{tabular}{ccccccccccccc}
\hline
\multirow{5}{*}{\small\textbf{EER}} &
\multicolumn{4}{c}{\textbf{GR}: 15.76} &
\multicolumn{4}{c}{\textbf{PH}: 15.98} &
\multicolumn{4}{c}{\textbf{EXT}: 14.63} \\
\cline{2-13}

&\multicolumn{6}{c}{\textbf{Studio recording}: 9.54} \vline & \multicolumn{6}{c}{\textbf{Home recording}: 19.96}  \\
\cline{2-13}

&\multicolumn{3}{c}{\textbf{Female}: 11.10} \vline & \multicolumn{3}{c}{\textbf{Male}: 7.53} \vline & \multicolumn{3}{c}{\textbf{Female}: 20.08} \vline & \multicolumn{3}{c}{\textbf{Male}: 19.78} \\
\cline{2-13}

 &\textbf{GR} & \textbf{PH} &\multicolumn{1}{c}{\textbf{EXT}} \vline&  \textbf{GR} & \textbf{PH} & \multicolumn{1}{c}{\textbf{EXT}} \vline& \textbf{GR} & \textbf{PH} & \multicolumn{1}{c}{\textbf{EXT}} \vline &  \textbf{GR} & \textbf{PH} & \textbf{EXT} \\
& 10.18 & 13.88 & \multicolumn{1}{c}{9.25} \vline& 8.33 & 7.14 & \multicolumn{1}{c}{7.14} \vline& 20.50 & 18.86 & \multicolumn{1}{c}{19.86}\vline & 20.83 & 19.78 & 18.74 \\ 
\hline 
&&&&&&&&&&&&\\
&&&&&&&&&&&&\\
\hline

\multirow{5}{*}{\small\textbf{WER}} &
 \multicolumn{4}{c}{\textbf{GR}: 26.00} &
\multicolumn{4}{c}{\textbf{PH}: 24.59} &
\multicolumn{4}{c}{\textbf{EXT}: 26.35}  \\
\cline{2-13}

&\multicolumn{6}{c}{\textbf{Studio recording}: 17.35} \vline & \multicolumn{6}{c}{\textbf{Home recording}: 31.96}  \\
\cline{2-13}

&\multicolumn{3}{c}{\textbf{Female}: 15.76} \vline & \multicolumn{3}{c}{\textbf{Male}: 19.39} \vline & \multicolumn{3}{c}{\textbf{Female}: 29.83} \vline & \multicolumn{3}{c}{\textbf{Male}: 35.44} \\
\cline{2-13}

 &\textbf{GR} & \textbf{PH} & \multicolumn{1}{c}{\textbf{EXT}} \vline  &  
 \textbf{GR} & \textbf{PH} & \multicolumn{1}{c}{\textbf{EXT}} \vline   & 
 \textbf{GR} & \textbf{PH} & \multicolumn{1}{c}{\textbf{EXT}} \vline   &  
 \textbf{GR} & \textbf{PH} & \multicolumn{1}{c}{\textbf{EXT}}  \\
&13.17 & 15.77 & \multicolumn{1}{c}{18.36} \vline& 21.64& 17.21& \multicolumn{1}{c}{19.32} \vline & 31.16 & 30.31 & \multicolumn{1}{c}{28.02} \vline& 35.86 & 31.69 & 38.77 \\ 

\botrule
\end{tabular}
\end{table}

The EER and WER values averaged across all the different evaluation criteria are shown in Table~\ref{tab:EER_WER_avg}. 
At first inspection of the results in Table~\ref{tab:EER_WER_avg} it can be noticed that the recording conditions have the most impact over the quality of the resulting voice both in terms of EER and WER. The studio recorded speakers exhibit on average a 50\% relative improvement over the EER and WER. With respect to the speaker's gender, for the studio recordings, the male voices performed better in terms of EER, and the female voices in terms of WER. For the home recordings, the female voices perform slightly better than the male voices in the WER evaluation. With respect to the text representation, the EXT transcription yields the best results for speaker similarity (EER), and the PH transcription for intelligibility (WER). But the differences between the systems are not statistically significant. This means that on average the text representation is not a determinant factor in the synthesis quality. 
The lowest WER (13.17\%) is achieved by the grapheme-based system with female voices.\footnote{We should note here that Romanian has rather simple letter-to-sound rules, and that these results may not be extended to more phonetically complex languages, such as English.} The fact that the \texttt{EXT} text representation does not perform better than the \texttt{PH} system could be interpreted as the fact that although the representation offers more information regarding the exact pronunciation of the text input, the network is not able to learn this additional correspondence.   

\begin{figure}[t!]
\scriptsize

\includegraphics[width=0.95\textwidth]{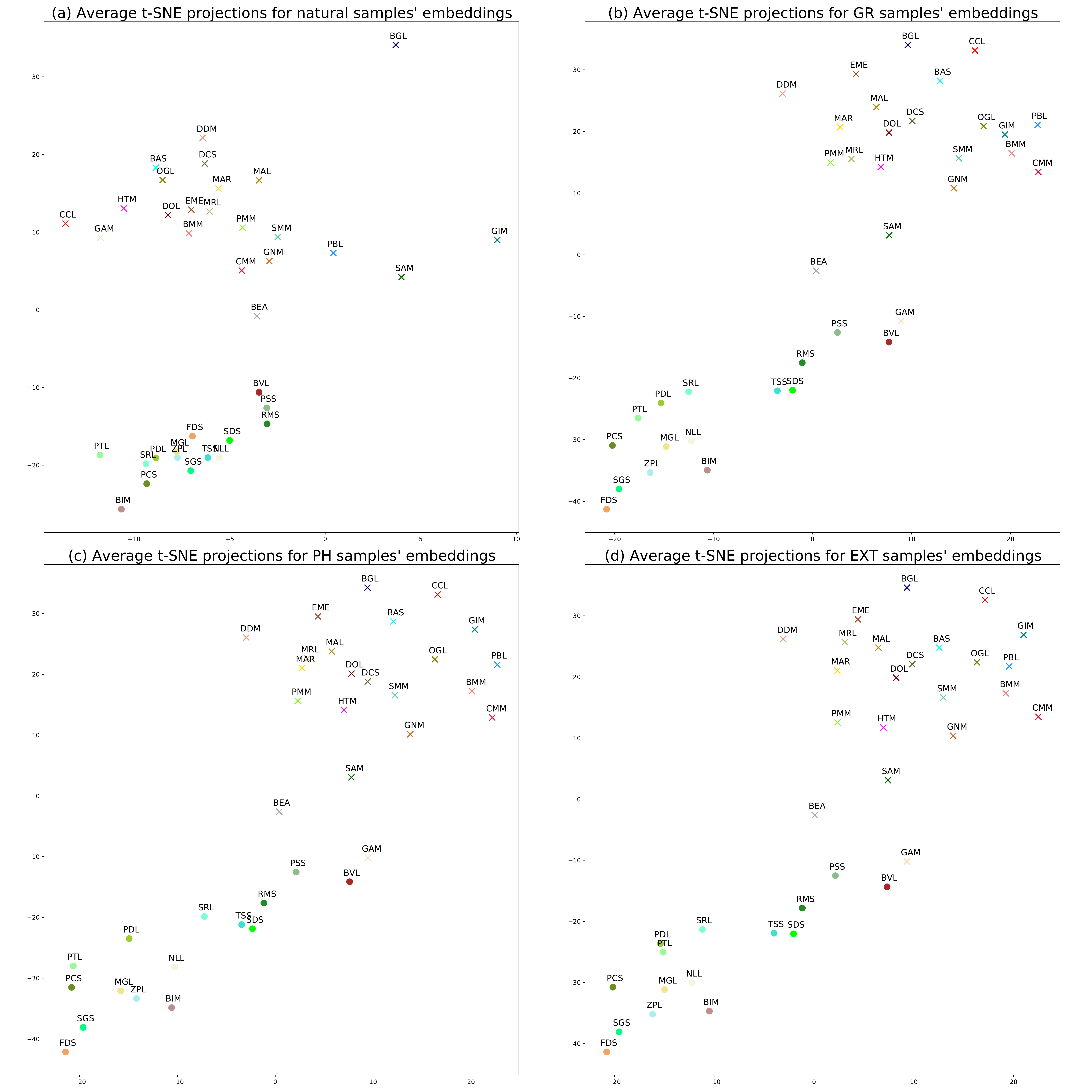}
\centering
\caption{t-SNE visualizations of the average speaker embedding for each speaker as obtained from the natural samples, and from the three TTS systems which use the different text representations (i.e. GR - graphemes, PH - phonemes, EXT - phonemes plus syllabification and lexical stress). The 'x' marks female speakers, and 'o' marks the male speakers.}
\label{fig:tsne_all2}
\end{figure}

\begin{figure}[h!]
\scriptsize
\includegraphics[width=0.6\textwidth]{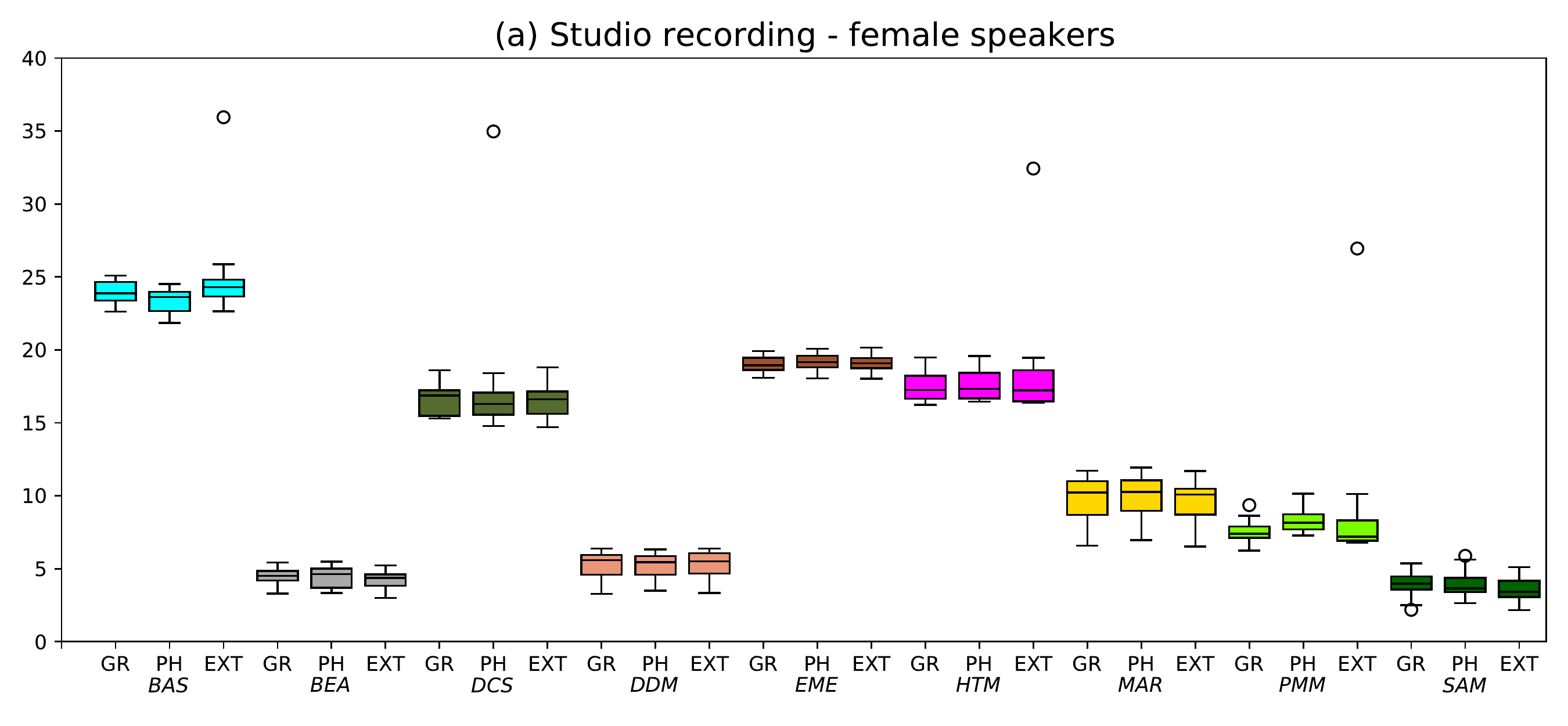}
\includegraphics[width=0.5\textwidth]{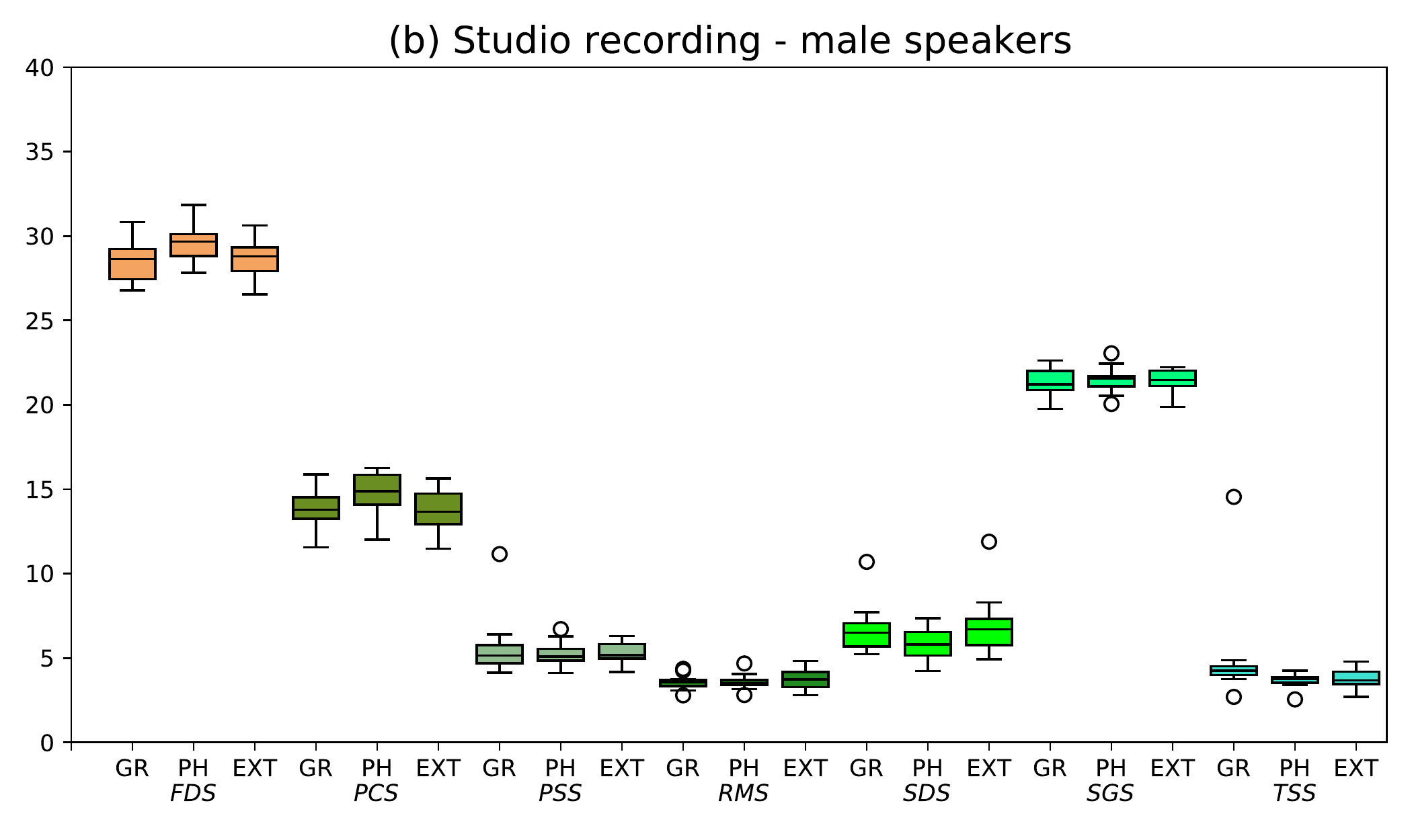}
\includegraphics[width=0.9\textwidth]{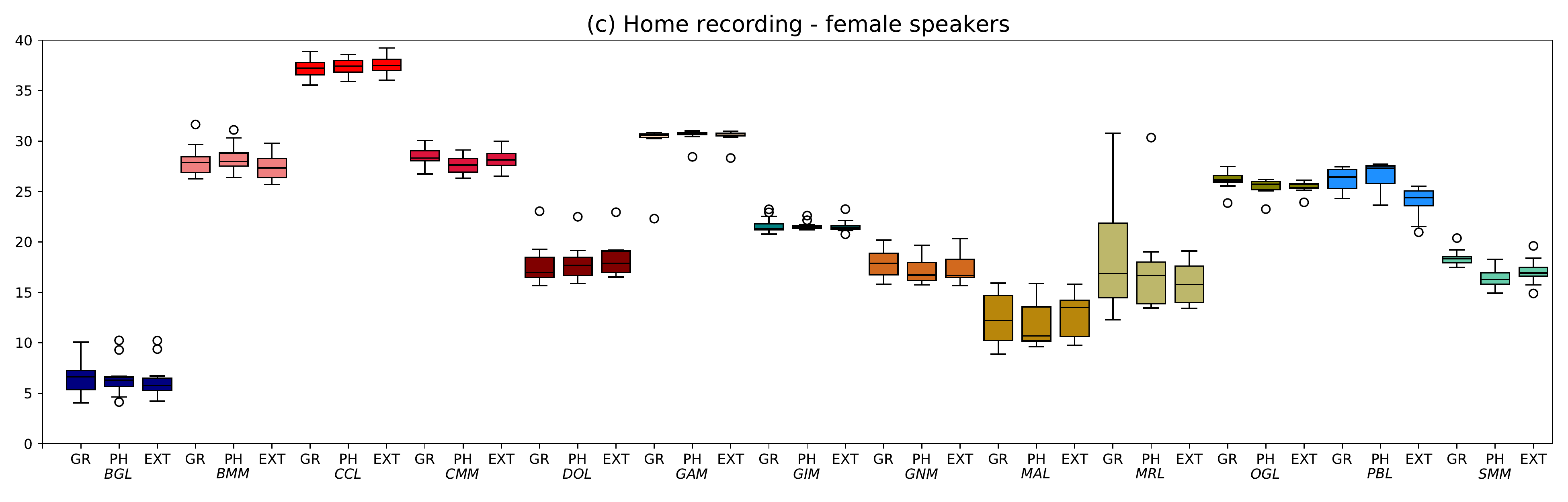}
\includegraphics[width=0.6\textwidth]{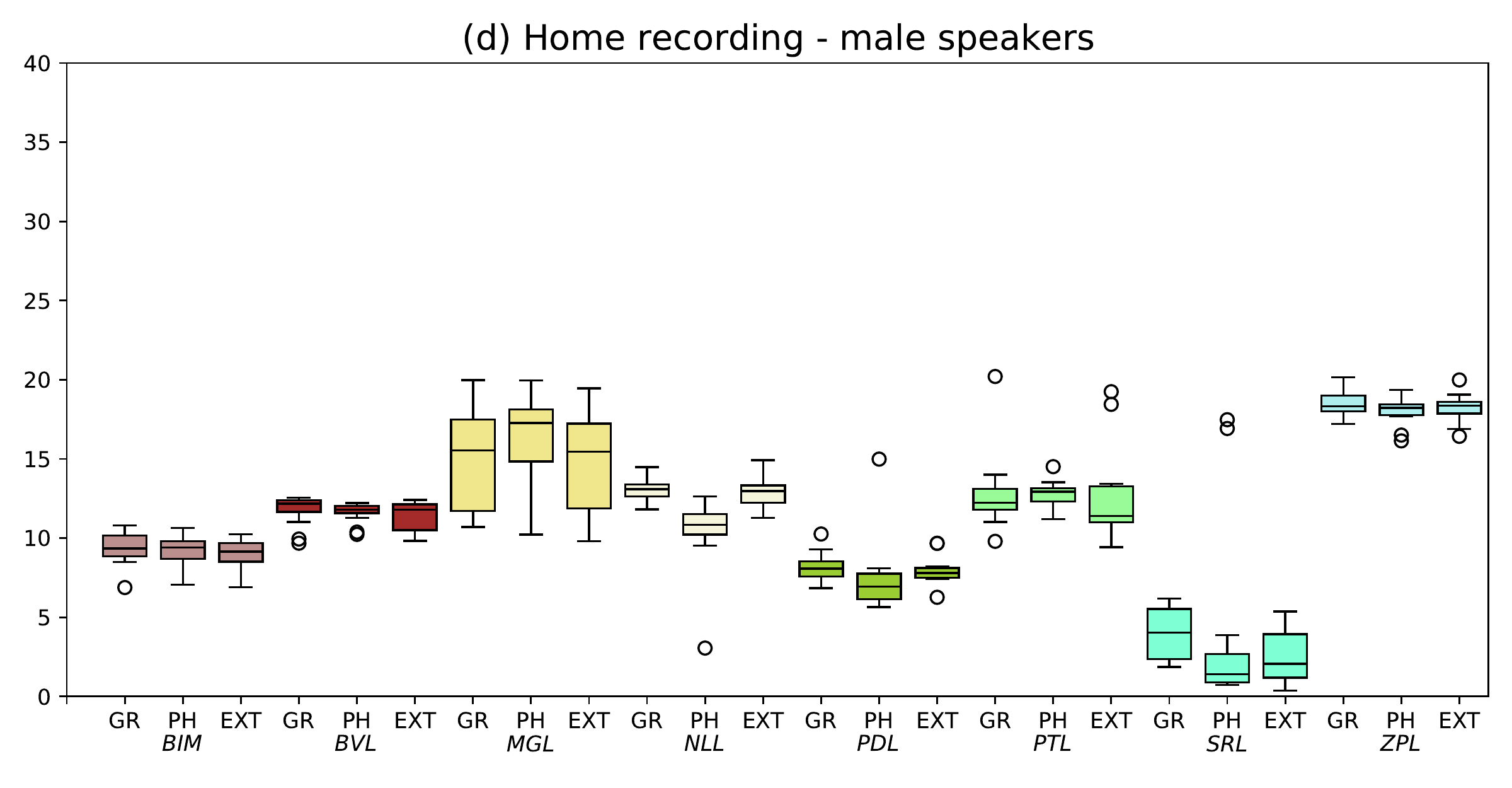}
\centering
\caption{Boxplots of the Euclidean distances calculated between the 2D t-SNE projections of the SV pairs of embeddings from natural and corresponding synthesised utterance. The plots are grouped by recording conditions, speaker gender and text input representations. Lower values are better.}
\label{fig:tsne_avg_record_cond_male_female}
\end{figure}

Breaking the above results at speaker level, we can notice some interesting facts. 
One speaker stands out in both the EER and WER evaluations, namely BVL. Its error rates are extremely high ($\overline{EER}:44.44\%$ , $\overline{WER}:90.66\%$) compared to all the other speakers. By inspecting its recordings more thoroughly, we discovered that the speaker consistently trimmed the beginning and ends of the utterances. This makes them unusable for the current purpose, or any other application where the correct speech-to-text alignment is required. Therefore, we removed the speaker from the SWARA 2.0 dataset. 

There are 3 voices which have perfect EER values: SAM, RMS and OGL across all text representation systems. RMS also performs best in terms of WER within its group (i.e. male speakers recorded in a studio environment), but it is not the case for SAM and OGL. Because the absolute error rates can be hard to interpret as is, we also examine the individual speaker embeddings.
For each speaker we passed the 12 natural and 36 synthesised test samples through the SV network and obtained the associated speaker embeddings. These embeddings were then projected into a 2D space using t-SNE~\cite{van2008visualizing}. 
Figure~\ref{fig:tsne_all2} plots the t-SNE visualisation of the average speaker embedding for each speaker as extracted from the natural or synthesised samples. The t-SNE projections were computed from all samples of all speakers and TTS systems so that the coordinates are directly comparable across the plots. It can be noticed that there is a rather clear separation between the male (marked with an 'o') and female (marked with a 'x') embeddings. Also, the embeddings of the synthesised samples (Figure~\ref{fig:tsne_all2}(b-d)) are slightly shifted from the natural ones (Figure~\ref{fig:tsne_all2}(a)), yet there are only minor shifts in between the synthesised samples' embeddings. To evaluate the shift of the synthesised samples versus the natural ones, we compute the Euclidean distance between each of the 12 natural test samples and their corresponding synthesised ones, and plot their statistics in Figure~\ref{fig:tsne_avg_record_cond_male_female}. The plots should be interpreted as how much does each speaker's synthesised output differ in terms of speaker identity with respect to its natural samples. The higher the values, the lower the speaker similarity. With very few exceptions, the distances between the natural and synthesised samples are consistent across the three TTS systems, and have rather tight distributions. For the studio recordings, there is a correlation between these distances and the EER values obtained by each of the 16 speakers. The Pearson correlation coefficient (PCC) for these values is 0.4. However, it is interesting to notice that this correlation is not preserved in the home recordings group of speakers (PCC=0.05). For example, speaker SRL has some of the lowest distances, yet its average EER and WER values are among the highest. On the other hand, speaker CCL exhibits very high distances, yet its EER is at most 8.33\% for all systems. Also, there is no correlation between the EER and WER values for the studio conditions (PCC=-0.01), but the home recordings' EER and WER are somewhat correlated, having a PCC of 0.35. 

\section{Conclusions}
\label{conclusions}

In this paper we performed an elaborate analysis and evaluation of the influence of speaker characteristics, recording conditions and text representations over the performance of a multi-speaker DNN-based text-to-speech synthesis system, namely the Tacotron2 architecture. We looked into the EER and WER values for each individual speaker, and plotted the distances between the embeddings of natural and synthesised samples as obtained from an accurate speaker verification network.

The results showed that a major role in the output of the synthesis is played by the availability of studio-quality recordings. All speakers which were recorded using their home setups had higher EER and WER rates. The text representation fed to the neural TTS system did not seem to influence the overall objective quality of the voices. With respect to the individual speakers there seem to be a few voice identities which are best suited for this TTS system, at least. It would be interesting to analyse their adequacy for other types of systems in order to draw definite conclusions regarding their spectral and prosodic characteristics. Another conclusion drawn from these experiments is the potential use of objective measures in order to asses the quality of the training data. Such as it was the case of the BVL speaker which at the first check seemed to have recorded the sentences correctly. Yet at the evaluation, the high EER and WER values revealed that the speaker consistently trimmed the start and end of the utterances. This deemed a set of recordings which were not truly aligned with the transcript, resulting in the erroneous synthesis output. 

Within this evaluation we did not perform any subjective listening tests, as the number of comparisons and samples needed to be evaluated would have made it unfeasible for a comfortable listener evaluation. However, we do plan to perform a series of further analyses of these results and select a subset of systems and speakers for a subjective evaluation.
As future work we would like to investigate even further how much does the voice identity truly influence the results of a TTS system: is it just a matter of chance, or are there any definite spectral and prosodic characteristics which influence the output. There is also the issue of augmenting the data of the less performing speakers, and how can a TTS system overcome the potential issues arising within the provided recordings.

\section*{Acknowledgements}

This work was supported by a grant of the Romanian Ministry of Research and Innovation, PCCDI – UEFISCDI, project number PN-III-P1-1.2-PCCDI-2017-0818/73, within PNCDI III.

\bibliographystyle{elsarticle-num}
\bibliography{main}

\end{document}